\documentclass[10pt,twocolumn]{article} 
\usepackage{simpleConference}
\usepackage{times}
\usepackage{graphicx}
\usepackage{amssymb}
\usepackage{url,hyperref}

\usepackage[table,xcdraw]{xcolor}
\usepackage{multirow}
\usepackage{fancyhdr}
\usepackage{lipsum} % For sample text

\fancypagestyle{firstpagefooter}{
    \fancyhf{}

    \fancyfoot[C]{© {Owner/Author | IEEE/OSA} . This is the author's version of the work. It is posted here for your personal use. Not for redistribution. The definitive Version of Record was accepted by {IEEE/OSA ECOC}.
    }
}

\begin{document}

%-------------------------------------------------- Title -----------------------------------------------------%

\title{A Pragmatic Power-Consumption Analysis for IPoWDM Networks with ZR/ZR+ Modules}%

%------------------------------------------------- Authors-----------------------------------------------------%

\author{Qiaolun Zhang$^{(1)^{*}}$, Annalisa Morea$^{(2)}$,  Massimo Tornatore$^{(1)}$\\
$^{(1)}$Politecnico di Milano, Corresponding author: $^{(*)}$qiaolun.zhang@mail.polimi.it  \\
$^{(2)}$Nokia, Vimercate, Italy
}

%\author{
%    Jiaheng Xiong\textsuperscript{(1)}, 
%    Qiaolun Zhang\textsuperscript{(1)}\textsuperscript{*}, 
%    Ruikun Wang\textsuperscript{(2)},
%    Alberto Gatto\textsuperscript{(1)}, 
%    Francesco Musumeci\textsuperscript{(1)}, \\
%    Massimo Tornatore\textsuperscript{(1)}\\
%    \textsuperscript{(1)} Politecnico di Milano, Milan, Italy, Corresponding author: *{\uline{qiaolun.zhang@mail.polimi.it}}
%   , \\
%   \textsuperscript{(2)} Beijing University of Posts and Telecommunications, Beijing, China 
%}

\maketitle

\thispagestyle{firstpagefooter}

\begin{abstract}
    We quantify and compare the power consumption of four IPoWDM transport network architectures employing ZR/ZR+ modules, considering different grooming, regeneration, and optical bypass capabilities. Results show that optical bypass is still the most power-efficient solution, reducing consumption by up to 30\%.
\end{abstract}
    
  %  50\% reduction in spectrum occupation compared to single-mode transmission and Full MIMO approach, and x\% reduction compared to dedicated path protection.

%-------------------------------------------------- Introduction Section -------------------------------------------------------%

\section{Introduction}
Network infrastructure power efficiency is a top priority for network operators that need to accommodate ever-growing traffic without increasing associated power-per-bit. The latest generation of ZR/ZR+ coherent transmission modules enables router client ports to support up to 400Gb/s using pluggable optics, which have lower cost and power consumption than traditional long-haul transponders~\cite{Wright2020,Zhu2021}. Meanwhile, power consumption of state-of-art IP routers is also decreasing rapidly~\cite{nokia21}. These emerging low-cost low-power silicon technologies have brought back in the foreground the \lq \lq opaque'' network architectures ~\cite{Musumeci2012}, where switching and regeneration of traffic are performed directly in IP routers, as their power consumption has significantly reduced  from 14.5W per Gb/s to 0.022 in the last decade~\cite{Musumeci2012,ciscoData2021}. %started nearing those of Reconfigurable Optical Add Drop Multiplexer (ROADM) technology used in \lq \lq transparent'' architectures.
% To investigate the renewed role of ZR/ZR+ modules on the power consumption of IPoWDM networks, in our study we compare four network architectures (all employing ZR/ZR+ modules) where switching and regeneration can be performed either (or both) at IP or at the optical layer. 
In this study, we want to investigate the possible interest of IPoWDM networks thanks to the use of ZR/ZR+ for four network architectures where switching and regeneration can be performed either (or both) at IP or the optical layer.

%: i) using IP routers to groom and regenerate traffic,  which merges several small channels into a large channel to reduce the network cost,  ii) using optical bypass to avoid unnecessary cost due to IP routers, iii) using ZR/ZR+ in back-to-back configuration to regenerate signal without consuming energy in IP routers.

% Existing studies mainly focus on verifying the reduction of cost of employing ZR/ZR+ compared with long-haul transponders~\cite{Wright2020,gumaste2022hardware,zami2022optimal,Gumaste2022}, while the evaluation of power consumption of IPoWDM architectures using ZR/ZR+ is not widely investigated.
% The only existing study on  power consumption where ZR/ZR+ are adopted is \cite{Gumaste2022}, which quantifies the power consumption focusing on the IP layer, without considering the impact of the node architecture on the number and type of devices to adopt on the IP and optical layers. The implications of ZR technology on power consumption when both IP and optical layer are considered are not modeled and discussed. 
Different from the work in the literature where ZR/ZR+ are compared with respect to long-haul transponders~\cite{Wright2020,gumaste2022hardware,zami2022optimal, Gumaste2022}, in this study we aim at evaluating the power consumption in an IPoWDM node depending on its architectures.
% Besides, studies on ZR/ZR+ on cost and energy consumption of IPoWDM networks~\cite{Melle2021,} do not thoroughly considered various network architectures with different settings for transparent network solutions.
% Ref.~\cite{Gumaste2022} measures the power consumption in the IP layer, excluding the components in the optical layer such as the energy consumption of ROADMs. 
%Moreover, various network architectures with different settings for transparent network solutions are not thoroughly considered.
At this scope, we define the number of used IP and optical devices in a node depending on its grooming, regeneration, and optical bypass capability. Then we provide a power model for the IPoWDM node and finally the power consumption of the various network architectures is estimated and compared.
% Thus, a study that extends the node power consumption model is needed, where the different options for an IPoWDM network architecture are thoroughly considered. In this work, we systematically classify different network architectures based on the different capabilities in terms of grooming, regeneration, and optical bypass. Then we provide an power model of network equipment at both IP and optical layer based on practical and up-to-date power consumption data. Finally we quantify the power consumption of the various network architectures under the proposed energy model.

\begin{figure*}[h]
   \centering
    \includegraphics[width=0.85\linewidth]{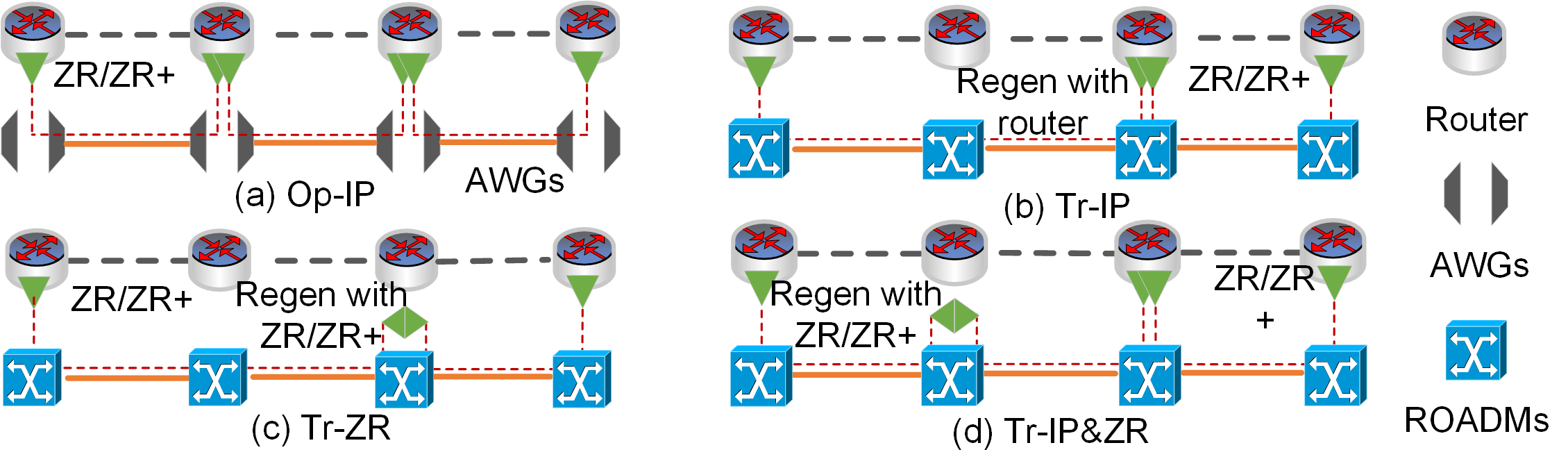}
    \caption{Network architectures for IPoWDM network.}
    \label{fig:network-architecture}
\end{figure*}
% \vspace{-16pt}

\section{IPoWDM Network Architectures}
Fig.~\ref{fig:network-architecture} summarizes the four considered IPoWDM network architectures. %to investigate the power consumption of the IPoWDM network employing ZR/ZR.
Fig.~\ref{fig:network-architecture} (a) depicts the \textbf{Opaque-IPoWDM (Op-IP)} case, where traffic is groomed and regenerated at each intermediate router. At the optical layer, we assume that lightpaths are muxed/demuxed using arrayed waveguide grating (AWG). Fig.~\ref{fig:network-architecture} (b) shows the \textbf {Transparent with regeneration at IP layer (Tr-IP)} case, where the traffic can be groomed and regenerated at intermediate routers (as in Op-IP), but nodes are also equipped with ROADMs that enable transparent lightpath bypass.  Fig.~\ref{fig:network-architecture} (c) shows the \textbf{Transparent with regeneration through ZR modules (Tr-ZR)} case, where to unload IP routers, we propose the introduction of regenerators realized by ZR/ZR+ in back-to-back configurations; with this configuration, no intermediate but only end-to-end grooming is allowed. Finally, Fig.~\ref{fig:network-architecture} (d) shows the \textbf{Transparent with IP and ZR regeneration (TR-IP\&ZR)} case, which combines the Tr-IP and Tr-ZR features, passing through the IP layer only when intermediate grooming is required.
% Although the Opaque-IPoWDMoWDM case does not have capabilities to bypass the node with ROADMs, it reduces the power consumption due to switching. 

\section{Power Consumption Model}
To compare different network architectures, we minimize the number of ZR/ZR+ modules and router ports (as a proxy for network cost); then, for the obtained solution we calculate the overall power consumption.%Note that we give priority to select ZR rather than ZR+ since the normalized cost of ZR and ZR+ is 1 and 2~\cite{Wright2020}, respectively.

Tab.~\ref{tab:normalized-power} reports the normalized power consumption of the network elements considered in our power model. In this work we only focus on the \textit{node} power consumption as it is the discriminator for the different studied architectures (the line remains the same). The IP router is composed of a fixed part (the chassis with fans, power pack, control board, and switching fabric), plus a modular part that depends on the total number of connected ZR/ZR+. When the regeneration is performed with 2 ZR/ZR+ in back-to-back configuration, they are not connected to the router.%Their respective power consumption is present in Table2. 

As for the optical node, it is modeled as follows. In the case of Op-IP, the optical node is composed of a terminal per direction realized by AWGs (Mux/Demux), each of which is associated with an optical amplifier (OA) to cope with line losses (2 OAs per node direction are required). Demultiplexed optical channels are then directly connected to ZR/ZR+ pluggables. Instead, for the Transparent cases, to allow the optical bypass functionality, WSSes are used as Mux/Demux to switch the optical channels either toward one of the node's output directions or toward ZR/ZR+ modules through an add/drop block (ADB). To improve overall power efficiency, a network element called I-ROADM~\cite{kundrat2019opening}, which integrates  WSSes and OAs for both directions (Go and Return), is used. Note that, due to the low power budget of the ZR module, low loss ADBs are mandatory (AWG is chosen in this study) and need to be associated to an OA.
An optical node requires also a number of shelves that must be dimensioned based on the number of I-ROADM, ADBs and OAs required in the node. The AWG is outside the shelf and hosted in the same rack. Our model accounts for the power consumption of the shelves, not of the rack.

\begin{table}[h]
\centering
\caption{Normalized power consumption of different IPoWDM network elements} 
\label{tab:normalized-power}
% \usepackage{multirow}
\begin{tabular}{|cc|c|}
\hline
\multicolumn{2}{|c|}{Device}                                                  & Power \\ \hline
\multicolumn{1}{|c|}{\multirow{4}{*}{\rotatebox{90}{IP side}}}      & ZR                      & 1     \\ \cline{2-3} 
\multicolumn{1}{|c|}{}                              & ZR+                     & 1.3   \\ \cline{2-3} 
\multicolumn{1}{|c|}{}                              & IP router fixed part           & 50    \\ \cline{2-3} 
\multicolumn{1}{|c|}{}                              & IP router modular part         & 4     \\ \hline
\multicolumn{1}{|c|}{\multirow{6}{*}{\rotatebox{90}{Optical part}}} & Shelf                   & 20    \\ \cline{2-3} 
\multicolumn{1}{|c|}{}                              & I-ROADM (bidirectional) & 4.1   \\ \cline{2-3} 
\multicolumn{1}{|c|}{}                              & OA (undirectional)      & 1.7   \\ \cline{2-3} 
\multicolumn{1}{|c|}{}                              & AWG                     & 0.3   \\ \cline{2-3} 
\multicolumn{1}{|c|}{} & \begin{tabular}[c]{@{}c@{}}Monitoring for opaque\\ (bidirectional)\end{tabular}      & 0.3 \\ \cline{2-3} 
\multicolumn{1}{|c|}{} & \begin{tabular}[c]{@{}c@{}}Monitoring for transparent\\ (bidirectional)\end{tabular} & 1.5 \\ \hline
\end{tabular}
\end{table}
% \vspace{-20pt}
\begin{figure}[h]
   \centering
        \includegraphics[width=\linewidth]{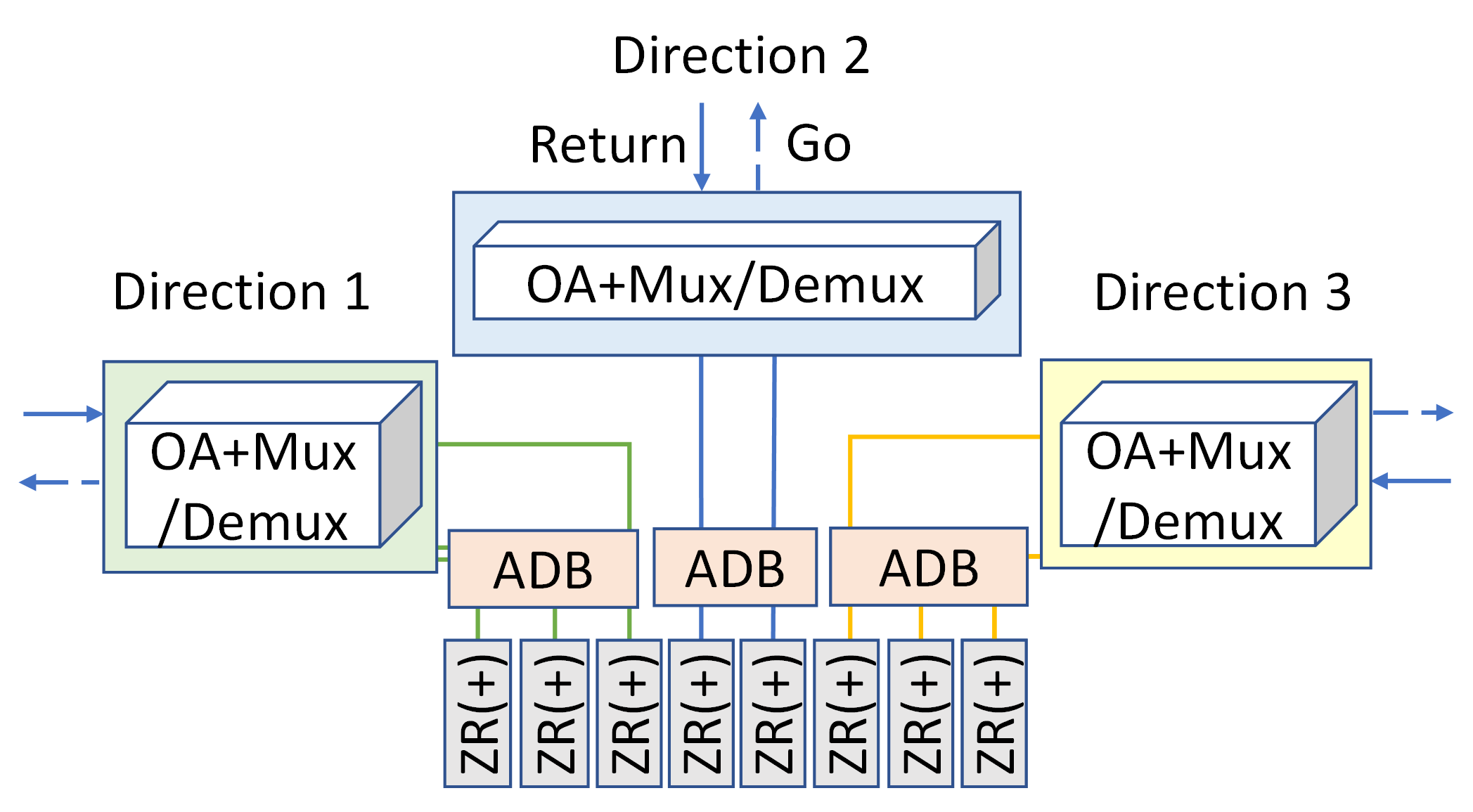}
    \caption{Optical node architecture \\[3.6pt]}
    \label{fig:oxc-architecture}
\end{figure}

\begin{table}[h]
\setlength\tabcolsep{1.6pt} 
   \centering
\caption{Modulation levels for ZR/ZR+ OEO devices} 
\label{tab:modulation-format}
\begin{tabular}{|c|c|c|c|}
\hline
Modules & MF    & Reach (km) & DR (Gb/s) \\ \hline
ZR      & 16QAM & 120        & 400       \\ \hline
ZR+     & 16QAM & 600        & 400       \\ \hline
ZR+     & 8QAM  & 1800       & 300       \\ \hline
ZR+     & QPSK  & 3000       & 200       \\ \hline
ZR+     & QPSK  & 3000       & 100       \\ \hline
\end{tabular}
\end{table}

\section{Routing, Modulation format, and Spectrum Assignment}
\begin{figure}[htb]
   \centering
        \includegraphics[width=0.95\linewidth]{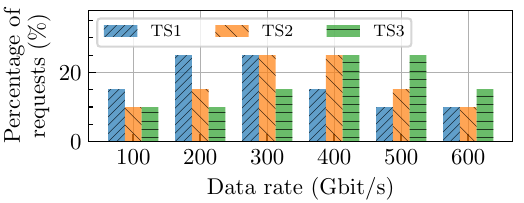}
    \caption{Percentage of requests (\%) for different TSs}
    \label{fig:traffic_percentage}
\end{figure}

\begin{figure*}[h]
   \centering
        \includegraphics[width=0.89\linewidth]{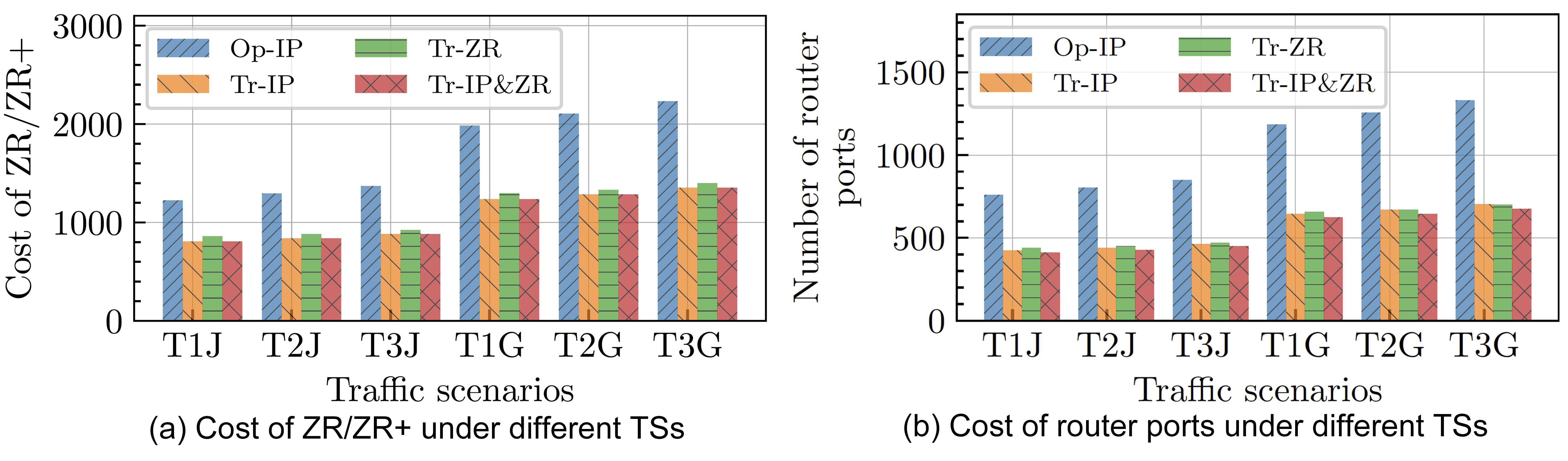}
    \caption{Cost of ZR/ZR+ and number of router ports for different TSs \\[3.5pt]}
    \label{fig:cost}
\end{figure*}

% \usepackage{multirow}
% \resizebox{\linewidth}{!}{%
\begin{table*}[]
\setlength\tabcolsep{1.68pt} 
   \centering
\caption{Power consumption for different IPoWDM network architectures on J14 and G17 topologies} 
\label{tab:result-power}
% \usepackage{multirow}
\begin{tabular}{|c|c|ccc|ccc|ccc|ccc|}
\hline
\rowcolor[HTML]{EFEFEF} 
\cellcolor[HTML]{EFEFEF} &
  \cellcolor[HTML]{EFEFEF} &
  \multicolumn{3}{c|}{\cellcolor[HTML]{EFEFEF}ZR/ZR+} &
  \multicolumn{3}{c|}{\cellcolor[HTML]{EFEFEF}IP router} &
  \multicolumn{3}{c|}{\cellcolor[HTML]{EFEFEF}Optical side} &
  \multicolumn{3}{c|}{\cellcolor[HTML]{EFEFEF}\textbf{Total power}} \\ \cline{3-14} 
\rowcolor[HTML]{EFEFEF} 
\multirow{-2}{*}{\cellcolor[HTML]{EFEFEF}Topology} &
  \multirow{-2}{*}{\cellcolor[HTML]{EFEFEF}\begin{tabular}[c]{@{}c@{}}Network\\ architectures\end{tabular}} &
  TS1 &
  TS2 &
  TS3 &
  TS1 &
  TS2 &
  TS3 &
  TS1 &
  TS2 &
  TS3 &
  TS1 &
  TS2 &
  TS3 \\ \hline
\cellcolor[HTML]{EFEFEF} &
  \cellcolor[HTML]{EFEFEF}Op-IP &
  \multicolumn{1}{c|}{899} &
  \multicolumn{1}{c|}{950} &
  1006 &
  \multicolumn{1}{c|}{2755} &
  \multicolumn{1}{c|}{2879} &
  3008 &
  \multicolumn{1}{c|}{464} &
  \multicolumn{1}{c|}{464} &
  464 &
  \multicolumn{1}{c|}{4120} &
  \multicolumn{1}{c|}{4295} &
  4479 \\ \cline{2-14} 
\cellcolor[HTML]{EFEFEF} &
  \cellcolor[HTML]{EFEFEF}Tr-IP &
  \multicolumn{1}{c|}{537} &
  \multicolumn{1}{c|}{558} &
  589 &
  \multicolumn{1}{c|}{1901} &
  \multicolumn{1}{c|}{1973} &
  2047 &
  \multicolumn{1}{c|}{722} &
  \multicolumn{1}{c|}{722} &
  722 &
  \multicolumn{1}{c|}{3162} &
  \multicolumn{1}{c|}{3255} &
  3359 \\ \cline{2-14} 
\cellcolor[HTML]{EFEFEF} &
  \cellcolor[HTML]{EFEFEF}Tr-ZR &
  \multicolumn{1}{c|}{573} &
  \multicolumn{1}{c|}{587} &
  615 &
  \multicolumn{1}{c|}{1862} &
  \multicolumn{1}{c|}{1932} &
  2004 &
  \multicolumn{1}{c|}{722} &
  \multicolumn{1}{c|}{722} &
  722 &
  \multicolumn{1}{c|}{3158} &
  \multicolumn{1}{c|}{3243} &
  3343 \\ \cline{2-14} 
\multirow{-4}{*}{\cellcolor[HTML]{EFEFEF}J14} &
  \cellcolor[HTML]{EFEFEF}Tr-IP\&ZR &
  \multicolumn{1}{c|}{537} &
  \multicolumn{1}{c|}{558} &
  589 &
  \multicolumn{1}{c|}{1878} &
  \multicolumn{1}{c|}{1946} &
  2018 &
  \multicolumn{1}{c|}{722} &
  \multicolumn{1}{c|}{722} &
  722 &
  \multicolumn{1}{c|}{3139} &
  \multicolumn{1}{c|}{3228} &
  3330 \\ \hline
\cellcolor[HTML]{EFEFEF} &
  \cellcolor[HTML]{EFEFEF}Op-IP &
  \multicolumn{1}{c|}{1424} &
  \multicolumn{1}{c|}{1510} &
  1600 &
  \multicolumn{1}{c|}{4028} &
  \multicolumn{1}{c|}{4224} &
  4427 &
  \multicolumn{1}{c|}{558} &
  \multicolumn{1}{c|}{558} &
  558 &
  \multicolumn{1}{c|}{6011} &
  \multicolumn{1}{c|}{6293} &
  6586 \\ \cline{2-14} 
\cellcolor[HTML]{EFEFEF} &
  \cellcolor[HTML]{EFEFEF}Tr-IP &
  \multicolumn{1}{c|}{821} &
  \multicolumn{1}{c|}{853} &
  896 &
  \multicolumn{1}{c|}{2631} &
  \multicolumn{1}{c|}{2745} &
  2858 &
  \multicolumn{1}{c|}{939} &
  \multicolumn{1}{c|}{939} &
  939 &
  \multicolumn{1}{c|}{4392} &
  \multicolumn{1}{c|}{4538} &
  4695 \\ \cline{2-14} 
\cellcolor[HTML]{EFEFEF} &
  \cellcolor[HTML]{EFEFEF}Tr-ZR &
  \multicolumn{1}{c|}{861} &
  \multicolumn{1}{c|}{885} &
  928 &
  \multicolumn{1}{c|}{2571} &
  \multicolumn{1}{c|}{2680} &
  2787 &
  \multicolumn{1}{c|}{939} &
  \multicolumn{1}{c|}{939} &
  939 &
  \multicolumn{1}{c|}{4373} &
  \multicolumn{1}{c|}{4504} &
  4655 \\ \cline{2-14} 
\multirow{-4}{*}{\cellcolor[HTML]{EFEFEF}G17} &
  \cellcolor[HTML]{EFEFEF}Tr-IP\&ZR &
  \multicolumn{1}{c|}{821} &
  \multicolumn{1}{c|}{853} &
  896 &
  \multicolumn{1}{c|}{2592} &
  \multicolumn{1}{c|}{2696} &
  2803 &
  \multicolumn{1}{c|}{939} &
  \multicolumn{1}{c|}{939} &
  939 &
  \multicolumn{1}{c|}{4353} &
  \multicolumn{1}{c|}{4489} &
  4640 \\ \hline
\end{tabular}
\end{table*}

Different from the traditional routing and spectrum (RSA) assignment problem, here we also need to select the type of transponders between ZR and ZR+ and determine its modulation format (MF). %The traditional RSA problem must evolve to a routing, modulation format, and spectrum assignment problem. 
Tab.~\ref{tab:modulation-format} listed the modulation levels, respective optical reaches, and the corresponding data rate (DR) for ZR/ZR+. Into the routing module, all channel are 100 GHz spaced.
%As shown in Tab.~\ref{tab:modulation-format}, 
% ZR has a fixed modulation format, which enables a data rate of 400 G/s. ZR+ has four different MFs, which may require different channel spacing and have different reach. %The traditional RSA problem must evolve to a routing, modulation format, and spectrum assignment problem.
%We assume that if one ZR/ZR+ is insufficient to serve a traffic request, one additional ZR/ZR+ will be used and traffic follows the same path, as in ~\cite{Zami2020}.

% In this work, we assume that each request must be served with one path. If the maximum data rate of a transponder is less than the required traffic, the network may require additional transponders. 
We developed an auxiliary graph-based heuristic algorithm, which is extensible for all the network architectures.
% The auxiliary link of the graph can be selected among pre-defined k-shortest path between its end nodes.%\footnote{The heuristic algorithm has been tested against an exact Integer Linear Program (ILP) model, showing an optimality gap below 5\% in most of the cases (tested over a 7-node topology).}.
% Detailed results regarding the benchmarking of the heuristics against ILP are not reported for lack of space. 
% The high-level structure of the heuristic is instead summarized as follows.
The high-level structure of the heuristic is summarized as follows.
We use the k-shortest path for routing of requests in the auxiliary graph, whose weight of auxiliary link is reduced if traffic can be groomed in it. Regarding to MF assignment, for Op-IP, Tr-IP, and TR-IP\&ZR, the algorithm selects the MF with the highest data rate, so that the residual capacity of the ligthpath can be used for grooming. 
For Tr-NoIP, the algorithm selects the MF requiring less spectrum. A first-fit policy is used for spectrum assignment.

\section{Case Studies and Results}
We perform our numerical evaluations on two topologies: 14-node Japan (J14) ~\cite{Ibrahimi2021} and 17-node Germany (G17) ~\cite{betker2003reference}. Each fiber can carry up to 50 100GHz-spaced channels.
%Main metric of the three topologies are shown in Table. 
Results are averaged over 10 instances of a full-mesh traffic matrix consisting of requests ranging from 100Gb/s and 600Gb/s with a 100Gb/s step.
We consider three traffic scenarios (TSs) in Fig.~\ref{fig:traffic-percentage}.
% We consider three traffic scenarios (TSs), which are summarized in Fig.~\ref{fig:traffic-percentage}. %The spectrum width used is 5 THz and all traffic requests are served in all cases.

% \begin{table}[]
% \centering
% \caption{Percentage of requests (\%) in traffic matrix for different TSs} 
% \label{tab:traffic-scenario}
% \begin{tabular}{|c|c|c|c|c|c|c|}
% \hline
% \begin{tabular}[c]{@{}c@{}}TR\\ (Gb/s)\end{tabular} & 100 & 200 & 300 & 400 & 500 & 600 \\ \hline
% TS1                                                 & 15  & 25  & 25  & 15  & 10  & 10  \\ \hline
% TS2                                                 & 10  & 15  & 25  & 25  & 15  & 10  \\ \hline
% TS3                                                 & 10  & 10  & 15  & 25  & 25  & 15  \\ \hline
% \end{tabular}
% \end{table}

% \usepackage{multirow}

%-------------------------------------------------- Conclusions Section ———————————————————————————%

\underline{\textbf{\textit{Cost Comparison.}}} 
% Tab.~\ref{tab:result-cost} 
Fig.~\ref{fig:cost} (a) and Fig.~\ref{fig:cost} (b) show the cost in terms of the number of ZR/ZR+ modules (weighted with the normalized cost of 1 and 2~\cite{Wright2020} for ZR and ZR+, respectively) and router ports used for different network architectures on J14 and G17 topologies.
% Note that the normalized cost of ZR and ZR+ is 1 and 2~\cite{Wright2020}, respectively.
The $TS$ of J14 and G17 is denoted as \textit{TkJ} and \textit{TkG}, respectively. As expected, in all cases, \emph{Tr-IP} achieves the smallest number of ZR/ZR+ and router ports, while \emph{Op-IP} has the largest number. By allowing grooming and regeneration, \emph{Tr-IP} reduces the number of ZR/ZR+ up to $36\%$ to $47\%$ compared to \emph{Op-IP} for J14 and G17.
%By allowing regeneration in both IP routers and using ZR/ZR+ in back-to-back configuration, 
\emph{Tr-IP\&ZR} decreases the cost 
% number of router ports with respect \emph{Tr-IP} 
by an additional $3\%$ ($4\%$) for J14 (G17). 
% Since Tr-ZR does not allow grooming, it requires up to 6\% and 5\% ZR/ZR+ than Tr-IP for J14 and G17.
% \todo[inline]{update the value}
% As for the number of ZR/ZR+ used, Opaque network solutions uses \% more ZR but uses only \% ZR+ than transparent network solutions. This is because opaque network solutions regenerate the traffic at all intermediate nodes and the traffic is likely to tr
% As shown in Tab.~\ref{tab:result-cost}, the Transparent-WithOrNoIP case is the one that with the lowest cost of ZR/ZR+ and consumes the least router ports for both J14 and G17. On the other hand, Opaque-IPoWDM is the one with the highest cost of ZR/ZR+. 

\underline{\textbf{\textit{Power consumption Comparison.}}} Tab.~\ref{tab:result-power} reports the power consumption for different network architectures. In both J14 and G17, \emph{Tr-IP\&ZR} has the lowest power consumption, while 
\emph{Op-IP} has the highest. Besides, power savings of \emph{Tr-ZR} wrt to \emph{Tr-IP} increase for traffic matrices with higher data rate, as large capacity requests have shorter reach and require more regenerators. In addition, due to optical bypass, \emph{Tr-IP} saves power consumption up to 29\% compared to \emph{Op-IP}.
% I moved down the comment so as to have the same layout than the previous part
% \todo[inline]{lot of re-writing to do, as discussion of results is too simplistic. The novelty introduced by ZR is not emphasized. it seems the same discussion that one would have without ZR}
% \todo[inline]{What about writing with ZR/ZR+, Opaque does not have too higher power consumption than Transparent case. And we can gain from the optical part with Opaque. Annalisa will update the value of Shelf tomorrow morning. But I think may not be so different from the current version.}

Let us now compare in more detail the power consumption of the various architectures.
% For J14, \emph{Tr-IP\&ZR} provides the highest total power consumption savings wrt Op-IP (around 25\% for all TSs). 
The power savings of Tr-IP\&ZR over Op-IP are mostly due to savings in power consumption of ZR/ZR+ and IP router ports. However, the lowest power consumption of IP routers is achieved with \emph{Tr-ZR} rather than \emph{Tr-IP\&ZR}, since \emph{Tr-ZR} does not consume any power in the intermediate nodes of a lightpath. Specifically, \emph{Tr-ZR} saves about 32\% of the power consumption of IP routers compared to \emph{Op-IP}. 
Moreover, due to grooming, \emph{Tr-IP} and \emph{Tr-IP\&ZR} have the lowest power consumption of ZR/ZR+, as both consume about 5\% less than \emph{Tr-ZR}. Similar considerations hold for G17 and J14.
% in terms of power consumption of ZR/ZR+.
% For G17, \emph{Tr-IP\&ZR} again achieves the highest power consumption savings, reaching about 30\% savings across all the three TSs wrt \emph{Op-IP} (i.e., the architecture with highest power consumption). 
One main difference between J14 and G17 is that the power consumption of the optical node in transparent architectures for G17 is about 68\% higher than in the opaque case, while it is about 55\% higher in J14. This is because the node degree in G17 is larger and it requires more shelves per node.
% Besides, \emph{Tr-ZR} saves about 36\% in  power consumption of IP routers compared with \emph{Op-IP}. 
% Moreover, grooming also saves energy consumption for \emph{Tr-IP} compared with \emph{Tr-ZR}, namely, 4.47\%, 3.50\%, and 3.39\%. \todo[inline]{Messages of results for G17 are exactly the same of J14, then why repeating the same thing? You should concentrate on differences}

\section{Conclusions}
Even in presence of the drastic reduction of power consumption of the optical routers and ZR/ZR+ seen in last years, the opaque network architecture is still more power-hungry than the transparent architectures employing optical bypass.
We demonstrate that using optical bypass saves up to 30\% of power consumption compared against opaque IPoWDM.

\bibliographystyle{abbrv}
\bibliography{refs}

%%%%%%%%%%%%%%%%%%%%%%%%%%%%%%%%%%%%%%%%%%%%%
%---------------------------------------------- End of Document -----------------------------------------------%
\end{document}